# How many radiographs are needed to re-train a deep learning system for object detection?

Raniere Silva

Khizar Hayat

Christopher M Riggs

Michael Doube

17 October 2022

## Abstract

### Background

Object detection in radiograph computer vision has largely benefited from progress in deep convolutional neural networks and can, for example, annotate a radiograph with a box around a knee joint or intervertebral disc. Is deep learning capable of detect small (less than 1% of the image) in radiographs? And how many radiographs do we need use when re-training a deep learning model?

### Methods

We annotated 396 radiographs of left and right carpi dorsal 75º medial to palmarolateral oblique (DMPLO) projection with the location of radius, proximal row of carpal bones, distal row of carpal bones, accessory carpal bone, first carpal bone (if present), and metacarpus (metacarpal II, III, and IV). The radiographs and respective annotations were splited into sets that were used to leave-one-out cross-validation of models created using transfer learn from YOLOv5s.

### Results

Models trained using 96 radiographs or more achieved precision, recall and mAP above 0.95, including for the first carpal bone, when trained for 32 epochs. The best model needed the double of epochs to learn to detect the first carpal bone compared with the other bones.

## Conclusions

Free and open source state of the art object detection models based on deep learning can be re-trained for radiograph computer vision applications with 100 radiographs and achieved precision, recall and mAP above 0.95.

## Introduction

Radiograph computer vision, radiograph automatic inspection and analysis by computer, has largely benefited from progress in machine learning methods, especially the ones based on deep convolutional neural networks, in many radiographic subspecialties including but not limited to chest imaging[1,2], musculoskeletal radiology[3–6], dental radiology[7], and pediatric radiology[8,9]. Progress has also be done in radiograph computer vision for veterinary medicine[10–12]. Two typical radiograph computer vision tasks are object classification (for example, annotating a radiograph with it's view[13] or the presence of a fracture[14]) and object detection (for example, annotating a radiograph with a box around a knee joint[15] or intervertebral disc[16]).

During the inception of deep convolutional neural networks models, creating new models required large data sets (> 100,000 images per class) and expensive computational resources[17]. For example, pioneering application of deep learning to radiographs used 256,458 radiographs to classify the presence or not of fracture and reported accuracy of 0.83 with a model based on the VGG architecture[14]. In 2022, many models are available for free under an open source license and can be fine-tuned (called transfer learning) for a task using a small data set (< 1,000 images per class) on an affordable office workstation running for a few hours. For example, deep learning models are able to classify elbow joint effusion with accuracy of 0.86 using 2,672 radiographs for training[18], classify failure in hip replacement with accuracy of 0.96 using 1,323 radiographs for training[19], detect a knee joint with precision of 0.99 using 8,634 radiographs[15], and detect an intervertebral disc with precision of 0.90 using 974 radiographs for training[16].

Despite the huge progress in deep learning, we still don't have the answer to "what is the smallest dataset that I need?" The reason that we don't have an answer is that the answer will change depending on many factors, including but not limited to the task that deep learning must solve, the architecture used, the model used for transfer learning, and the data augmentation pipeline that might include the generation of synthetic data. On classification of radiographs, it's reported the need of 100 images per class in the training dataset to achieve average accuracy of 0.89 based on experiments with the GoogLeNet architecture implemented in Caffe[20]. Other studies reinforce the recommendation of at least 100 images per class[13]. Latest report by Google Health indicates that it is possible to achieve an area under the receiver operating characteristic curve (AUC) of 0.95 in classifying microbiology-confirmed tuberculosis with only 45 images[21]. On object detection in radiographs, no previous study was conducted to answer the smallest dataset needed. Investigation toward the smallest dataset needed for detection of car and person in street scene images reports 14,962 images to achieve precision of 0.39 for car and 0.62 for person[22]. Translate results from street scene[22] to radiographs is hard because of the differences between the two type of images. In this paper, we try to answer "what is the smallest training dataset of a single radiograph view from a homogeneous population required to produce an object detection model with average accuracy above 0.95?"

## Materials and Methods

### Deep Convolutional Neural Networks Selection

Deep convolutional neural networks architectures for object detection are based in R-CNN[23] or YOLO[24]. We selected YOLOv5[25], an architecture based on YOLO, because YOLO is faster than R-CNN[24,26], showed promising results in other radiograph computer vision studies[27], and it's available under a open source license. Among YOLOv5 architecture family, we selected YOLOv5s because of it's "small" number of parameters (7.2 millions) of the model that translate in a faster detection time.

### Data Acquisition

We collected pre-sale radiograph sets of racehorses from The Hong Kong Jockey Club as mentioned in a previous study[13]. The total number of pre-sale radiograph sets available for this study was 198. For each pre-sale radiograph set, we extracted both the left and right carpi dorsal 75º medial to palmarolateral oblique (DMPLO) projection. (Figure 3.1).

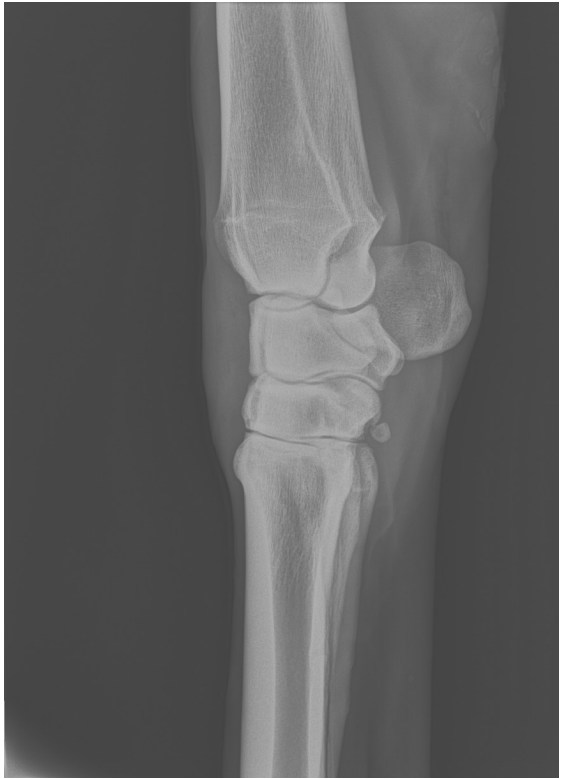

*With first carpal bone.*

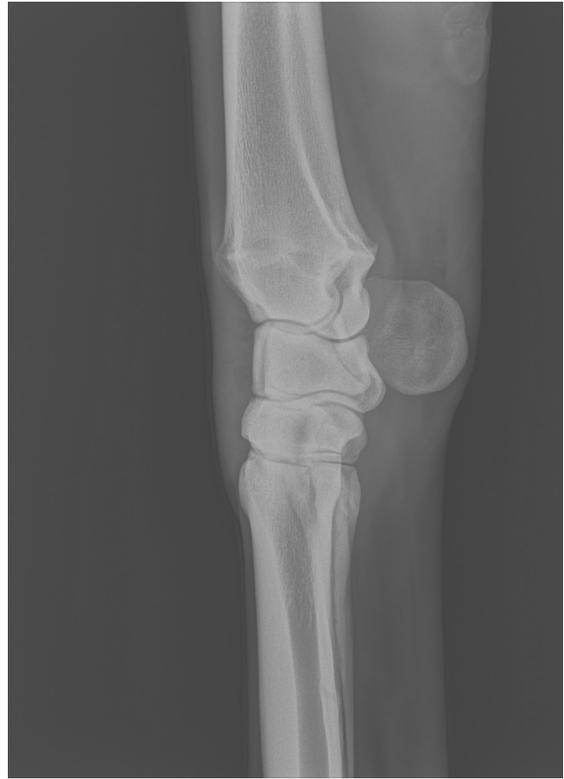

*Without first carpal bone.*

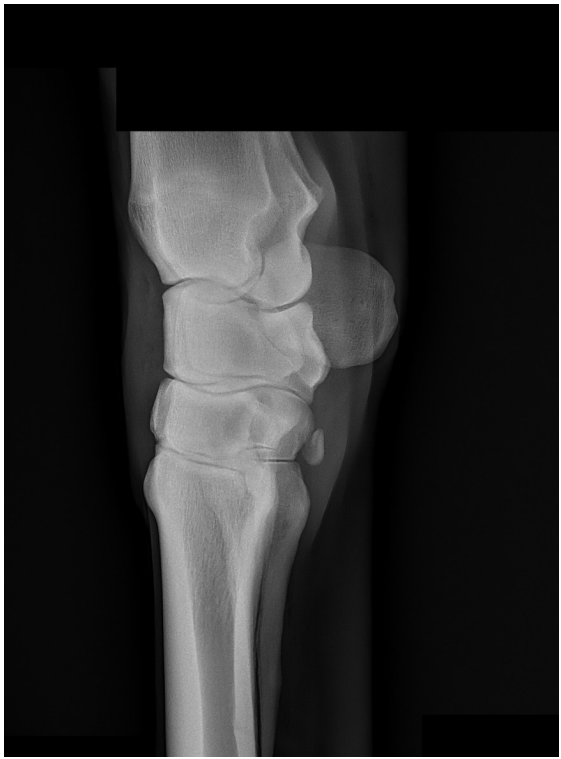

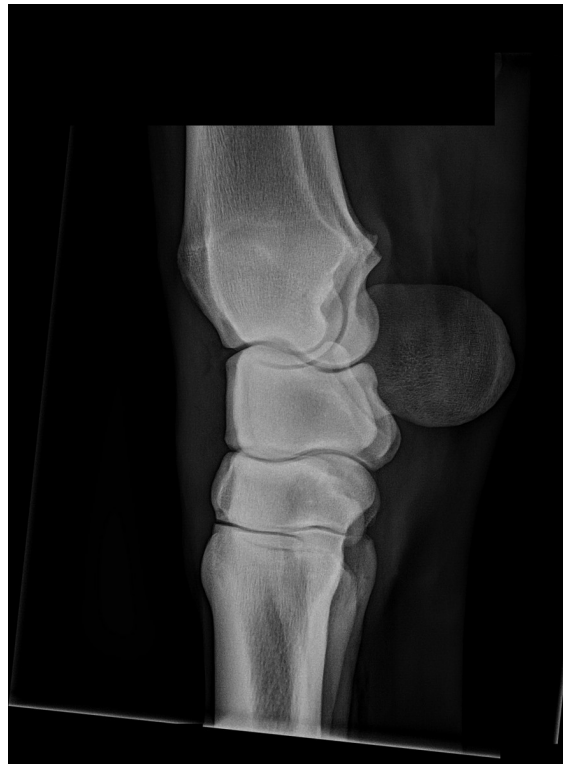

*With first carpal bone and text burned in cover by black rectangle.*   *Without first carpal bone and text burned in cover by black rectangle.*

*Figure 3.1: Illustration of left carpus dorsal 75º medial to palmarolateral oblique radiographs collected from the The Hong Kong Jockey Club.*

The carpus DMPLO radiograph was selected for this study because it represents a complex anatomical structure with multiple different bones superimposed or juxtaposed in a single joint and one of the bones visible on this projection, the first carpal bone, is variably present (it is a vestigal remnant present in approximately 25% of horses[28,29]). Furthermore, the carpus DMPLO radiograph was selected because it is used in the treatment of fracture of the accessory carpal bone where computer-assisted orthopaedic surgery was recently introduced[30].

## Data Annotation

Carpus DMPLO radiographs in the original resolution (Figure 3.2) were annotated with the location of radius, proximal row of carpal bones, distal row of carpal bones, accessory carpal bone, first carpal bone (if present), and metacarpus (metacarpal II, III, and IV) in the format of bounding boxes (Figure 3.3) by the first author using a self hosted instance of Label Studio v1.4.1 (Heartex). The bounding boxes were manually reviewed by the second author comparing the bounding boxes to the radiographs using the same a self hosted instance of Label Studio. After review, all bounding boxes were exported into YOLO Darknet format[24] using Label Studio. The first carpal bone was present in 26% of the racehorses participating in this study and the average bounding boxes size of the first carpal bone was 0.1% of the original image (Table 3.1).

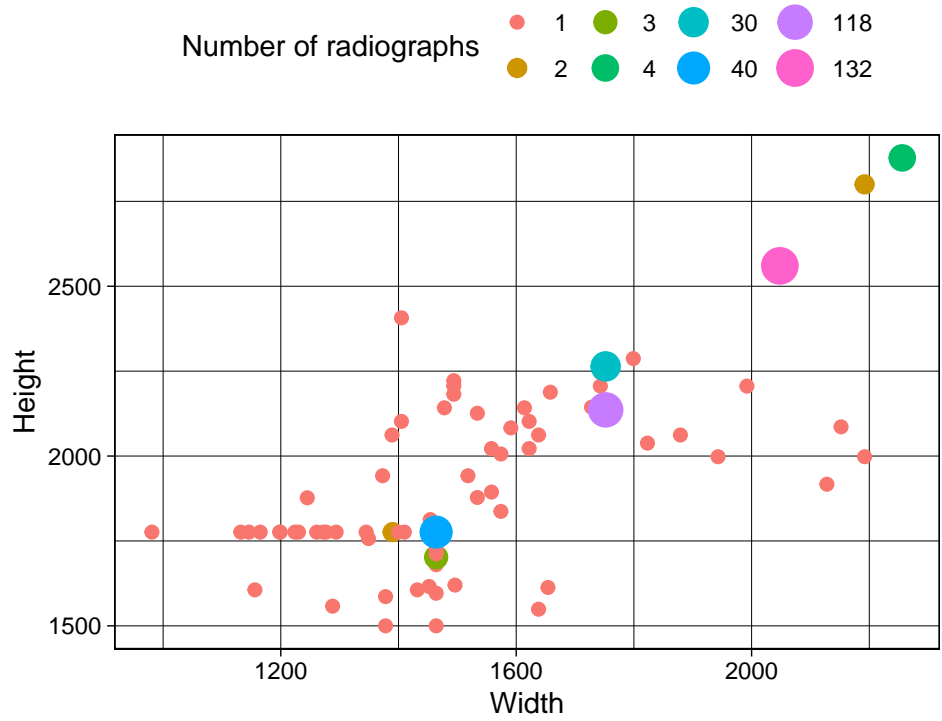

*Figure 3.2: Distribution of radiographs original resolution.*

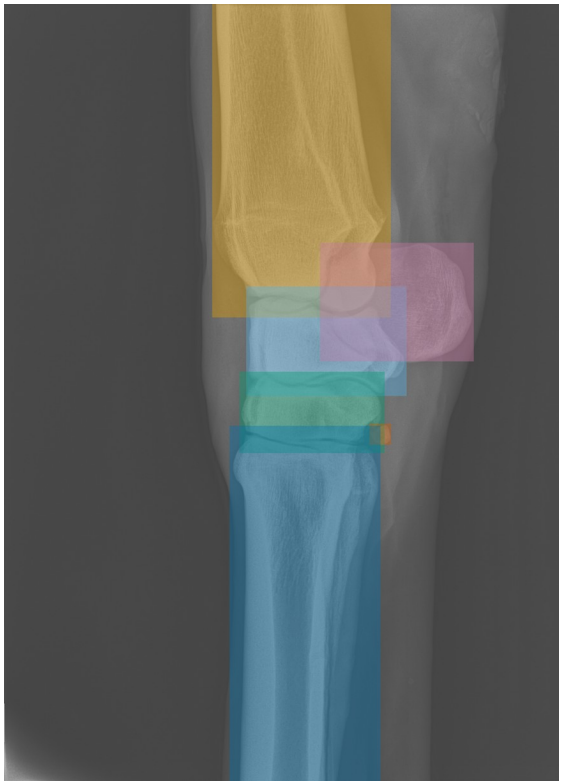

*With first carpal bone.*

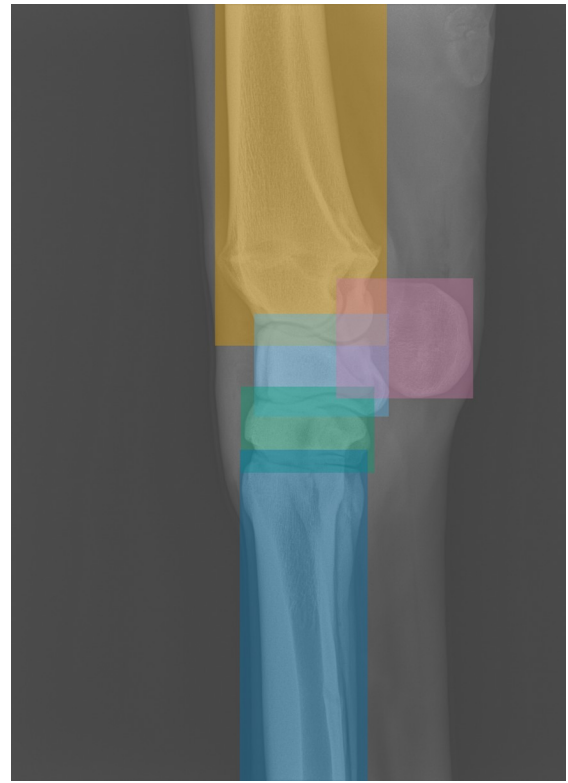

*Without first carpal bone.*

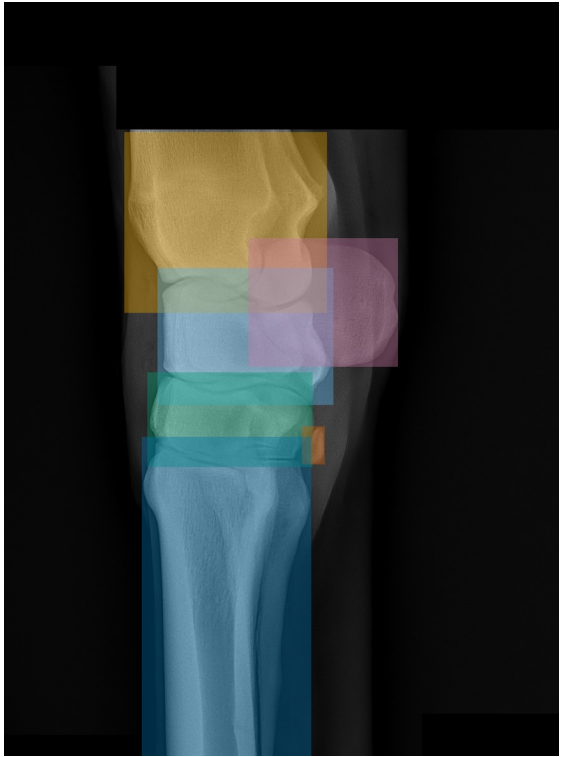 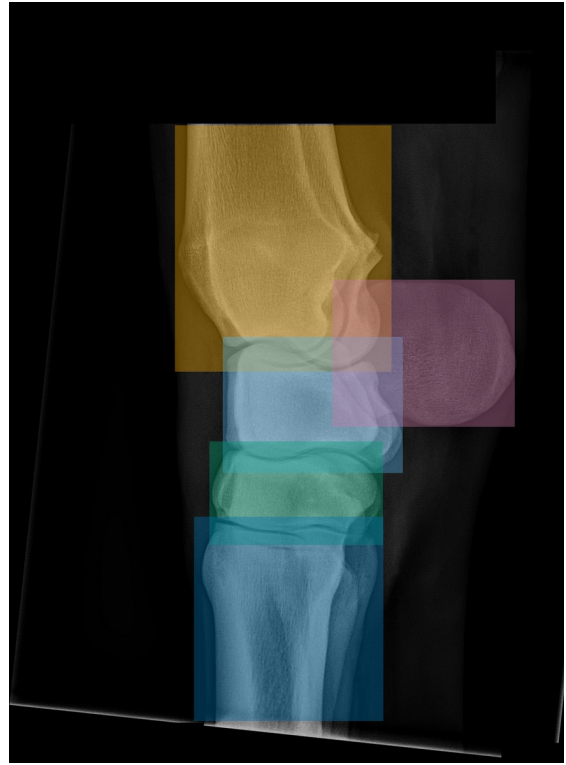

*With first carpal bone and text burned in cover by black rectangle.*    *Without first carpal bone and text burned in cover by black rectangle.*

Figure 3.3: Illustration of bounding boxes of left carpus dorsal 75º medial to palmarolateral oblique radiographs.

Table 3.1: Descriptive statistics of bounding boxes area in the original resolution and low resolution used by YOLOv5s (*).

| Bone | Min | Mean | Median | Max | Min (*) | Mean (*) | Median (*) | Max (*) |
|---|---|---|---|---|---|---|---|---|
| radius | 99157 | 488062 | 488502 | 980928 | 19955 | 49850 | 49898 | 80466 |
| acessory carpal | 85320 | 167618 | 174937 | 261016 | 11335 | 17589 | 16606 | 35900 |
| proximal carpal row | 87360 | 155892 | 156441 | 288325 | 10604 | 16571 | 15528 | 37120 |
| distal carpal row | 65650 | 114628 | 114415 | 219816 | 7943 | 12176 | 11532 | 27501 |
| first carpal bone | 1088 | 5166 | 5254 | 11781 | 172 | 531 | 503 | 1355 |

| Bone | Min | Mean | Median | Max | Min (*) | Mean (*) | Median (*) | Max (*) |
|---|---|---|---|---|---|---|---|---|
| metacarpal | 9711 | 417293 | 411240 | 862257 | 1025 | 42773 | 42117 | 104511 |

## Data Split

Radiographs were split into 6 data sets using scikit-learn (version 1.0.2) such that the data set's were balanced regarding the number of radiographs with the first carpal bone present or not (Figure 3.4).

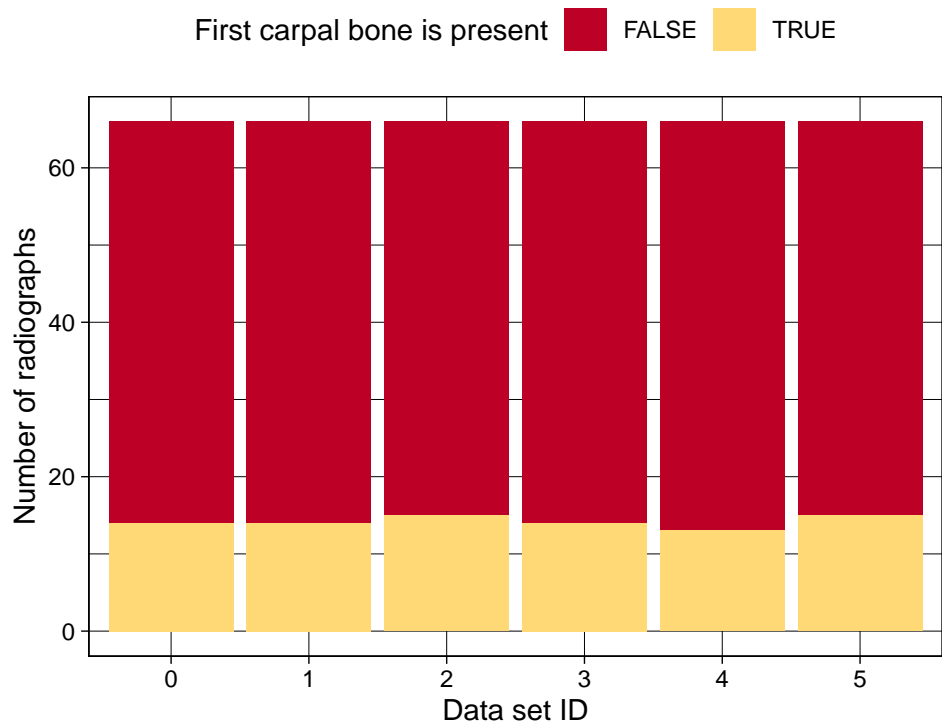

Figure 3.4: Distribution of radiographs across datasets.

One of the data sets were named test (data set ID = 5) and the other 5 sets were used for leave-one-out cross-validation (identified by the data set ID of the left out data set). Each one of the 5 sets for training (left out data set ID = 1-4) were used to generate subsets with different number of radiographs (Table 3.2), full details of the protocol in figshare doi:10.6084/m9.figshare.21061951. A Python script was used to create the data for each experiment in the YOLO format.

*Table 3.2: Composition of Secondary Dataset.*

| Secondary Dataset ID | Radiographs with first carpal bone | Radiographs with first carpal bone | Total number of radiographs |
|---|---|---|---|
| 1 | 8 | 8 | 16 |
| 2 | 16 | 16 | 32 |
| 3 | 24 | 24 | 48 |
| 4 | 32 | 32 | 64 |
| 5 | 40 | 40 | 80 |
| 6 | 48 | 48 | 96 |
| 7 | 56 | 56 | 112 |
| 8 | 64 | 64 | 128 |
| 9 | 64 | 80 | 144 |
| 10 | 64 | 96 | 160 |
| 11 | 64 | 112 | 176 |
| 12 | 64 | 128 | 192 |
| 13 | 64 | 144 | 208 |
| 14 | 64 | 160 | 224 |
| 15 | 64 | 176 | 240 |
| 16 | 64 | 192 | 256 |

## Training

Training based on leave-one-out cross-validation was conducted for 32 epochs using YOLOv5s architecture[25] and started with a model pretrained on the COCO dataset[31] (version 6.2). YOLOv5s used RGB $640 \times 640$ pixels image as input and batch size of 16. YOLOv5s uses data augmentation and combines parts of multiple images to create new ones (Figure 3.5).

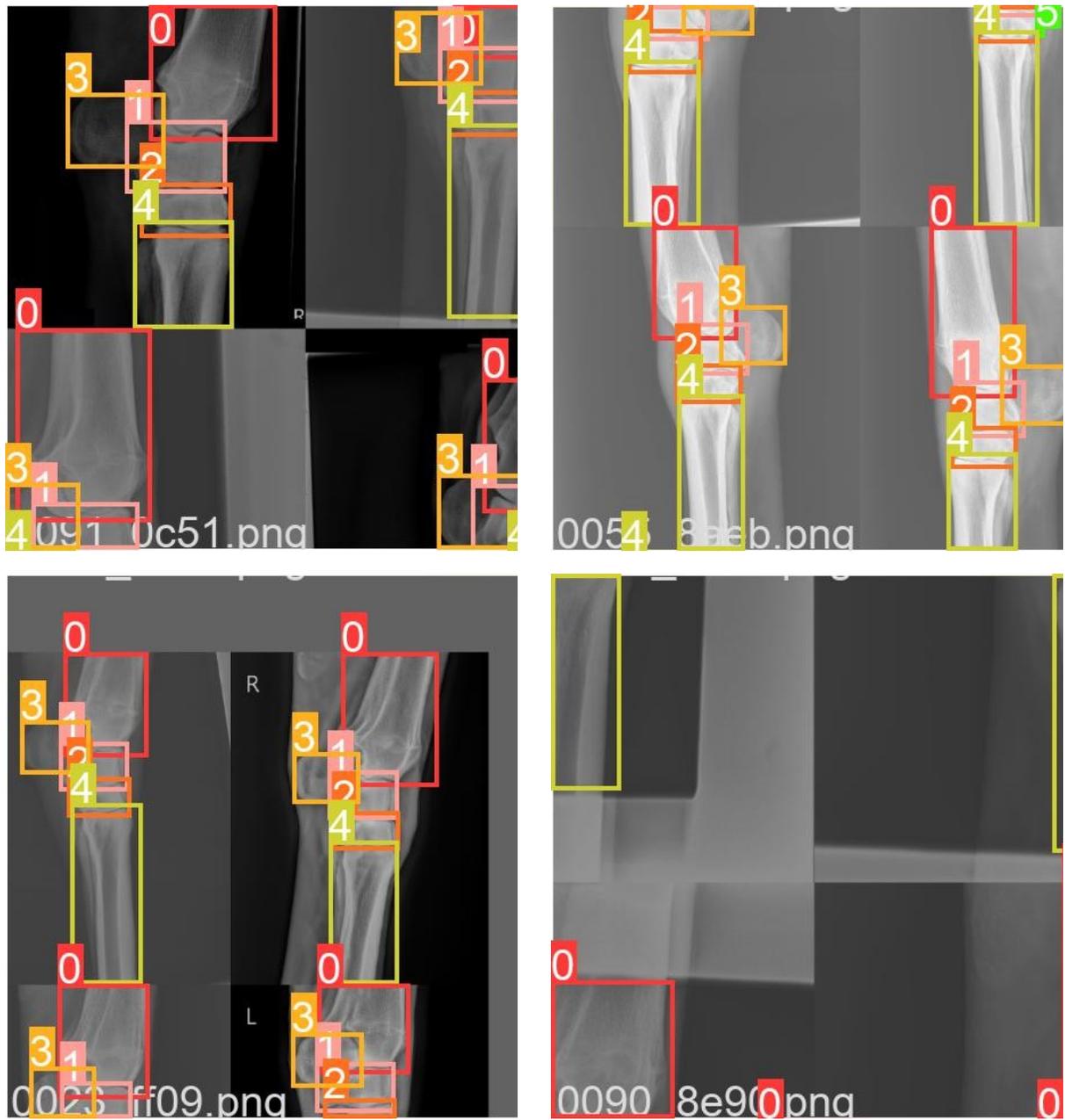

*Figure 3.5: Illustration of data augmentation used during training. Bounding boxes are included in the illustration for reference to the reader.*

At the end of each epoch, metrics (precision, recall, and mean average precision (mAP)[32,33]) were calculated using the validation data set (the one left out) and stored for further analysis. For calculation of the metrics, all bounding boxes calculated by the model are classified into true positive, true negative, false positive, or false negative using bounding box evaluation[32,33].

Computation was performed on a Precision 5820 workstation (Dell Hong Kong) equipped with an Intel(R) Xeon(R) W-2123 3.60GHz CPU, 64GiB (4 × 16GiB DDR4 2666MHz) of RAM, and NVIDIA Quadro RTX 4000 GPU running Ubuntu 21.10 (Linux 5.13), Python (version 3.9.9), NumPy (version 1.21.2), PyTorch (version 1.10.2), and CUDA (version 11.3).

## Testing

The best model of each leave-one-out cross-validation iteration was tested on the Precision 5820 workstation (Dell Hong Kong) described previously using the test dataset and metrics (precision, recall, and mAP0.5) were calculated and stored for further analysis.

## Analysis

Analysis of all metrics collected was conducted using descriptive statistics and data visualisation computed with R version 4.1.3 (2022-03-10) on the Precision 5820 workstation (Dell Hong Kong) described previously.

## Results

The best model of each leave-one-out cross-validation iteration was tested using the test data set the model never saw before and metrics were collected. Metrics improved with the increase of the number of radiographs in the training data set (Figure 4.1, Figure 4.2, Figure 4.3). With 96 or more radiographs, some models of the leave-one-out cross-validation achieved precision, recall and mAP above 0.95 (Figure 4.1, Figure 4.2, Figure 4.3). Unbalanced training data set (with more than 128 radiographs) showed metrics similar to the largest balanced training data set (Figure 4.1, Figure 4.2, Figure 4.3).

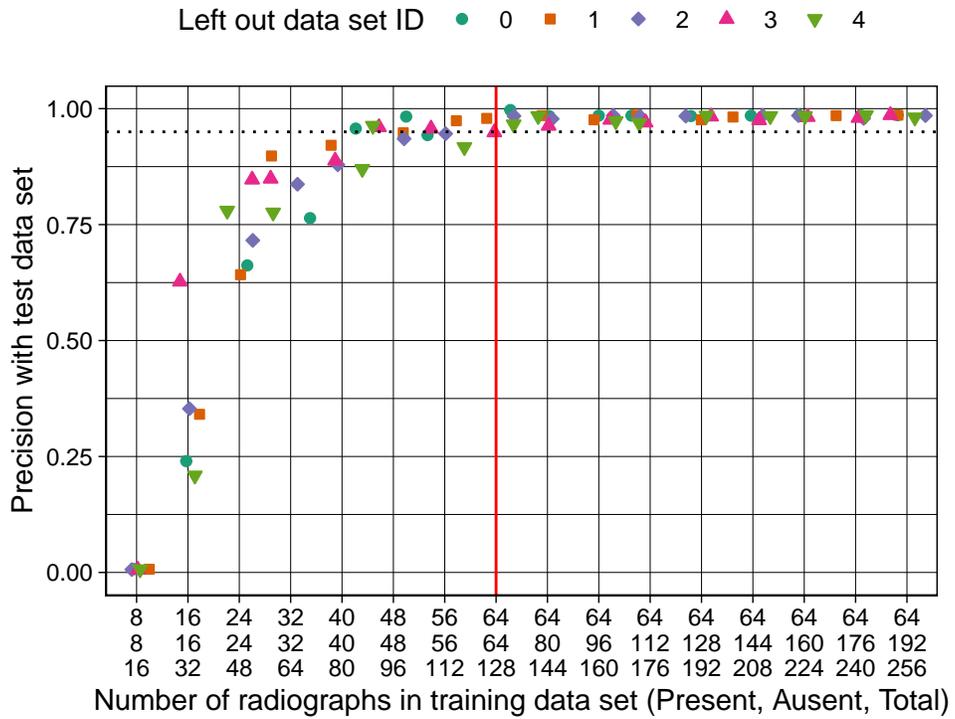

*Figure 4.1: Evolution of precision calculated using test data set as function of number of radiographs used for training. Data set is balanced at the left of vertical red line and unbalanced at the right of the vertical red line. Dotted horizontal line indicate 0.95.*

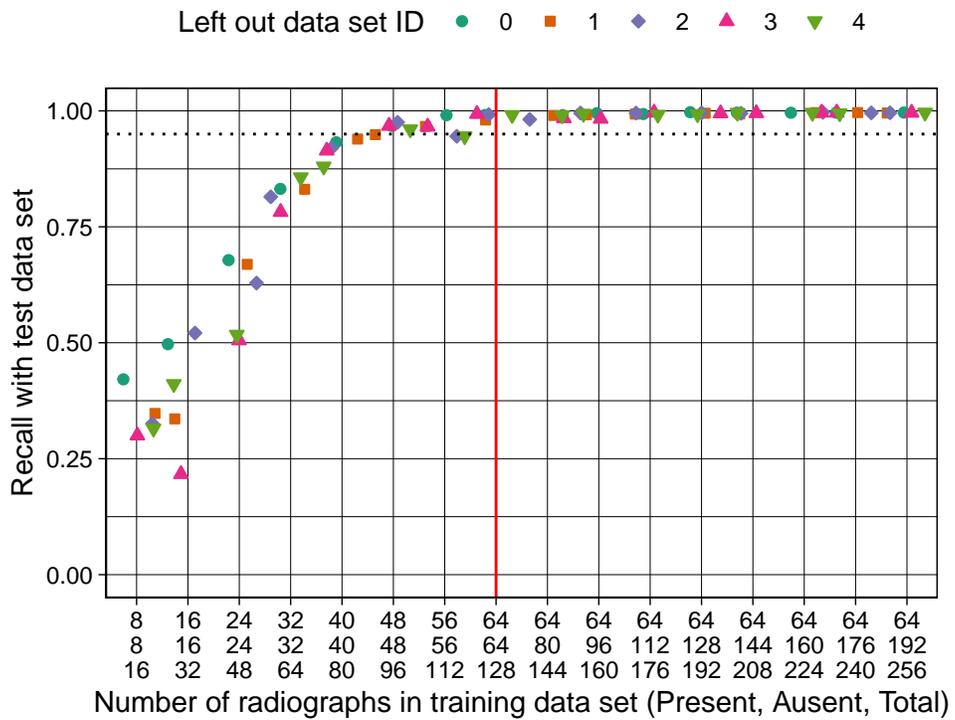

*Figure 4.2: Evolution of recall calculated using test data set as function of number of radiographs used for training. Data set is balanced at the left of vertical red line and*

*unbalanced at the right of the vertical red line. Dotted horizontal line indicate 0.95.*

Figure 4.3: Evolution of mAP calculated using test data set as function of number of radiographs used for training. Data set is balanced at the left of vertical red line and unbalanced at the right of the vertical red line. Dotted horizontal line indicate 0.95.

The best model (left out data set ID = 1 with 240 radiographs) achieved precision of 0.985, recall of 0.996 and mAP of 0.99. Table 4.1 has metrics of follow up models.

Table 4.1: Best models based on mAP for all classes.

| Left out data set ID | Number of Radiographs for training | Precision | Recall | mAP |
|---|---|---|---|---|
| 1 | 240 | 0.985 | 0.996 | 0.990 |
| 3 | 256 | 0.986 | 0.996 | 0.986 |
| 2 | 256 | 0.985 | 0.996 | 0.984 |

## Data Visualisation of Epoch

Increase the number of epochs improved the metrics during the first 20 epochs and maintained metrics with more than 20 epochs (Figure 4.4, Figure 4.5, Figure 4.6).

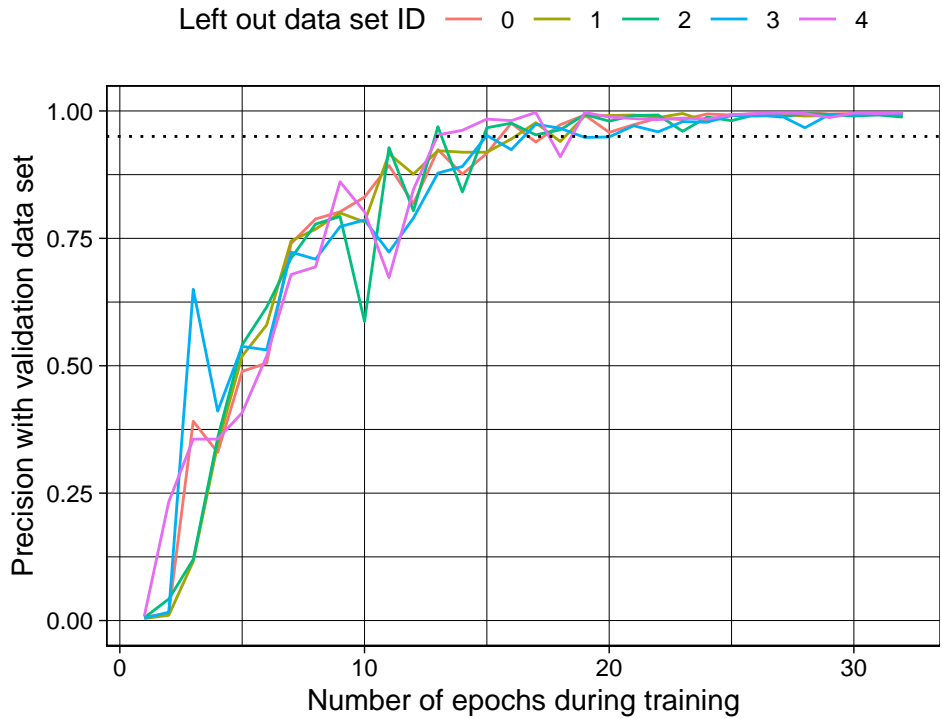

*Figure 4.4: Evolution of precision calculated using test data set as function of number of epochs during training with 128 radiographs in the training data set. Dotted horizontal line indicate 0.95.*

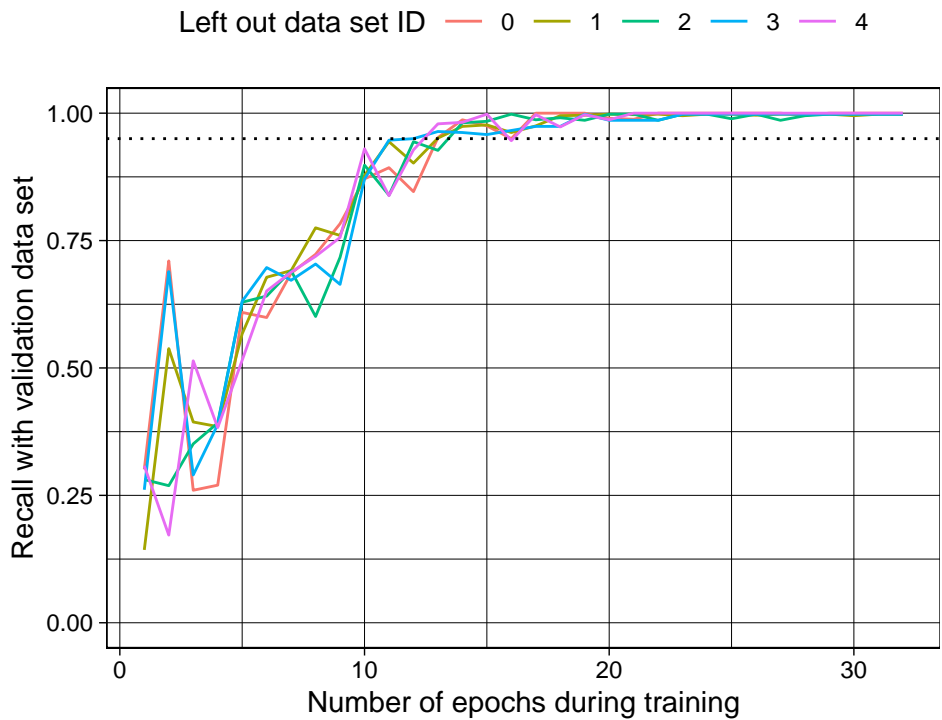

*Figure 4.5: Evolution of recall calculated using test data set as function of number of epochs during training with 128 radiographs in the training data set. Dotted horizontal line indicate*

0.95.

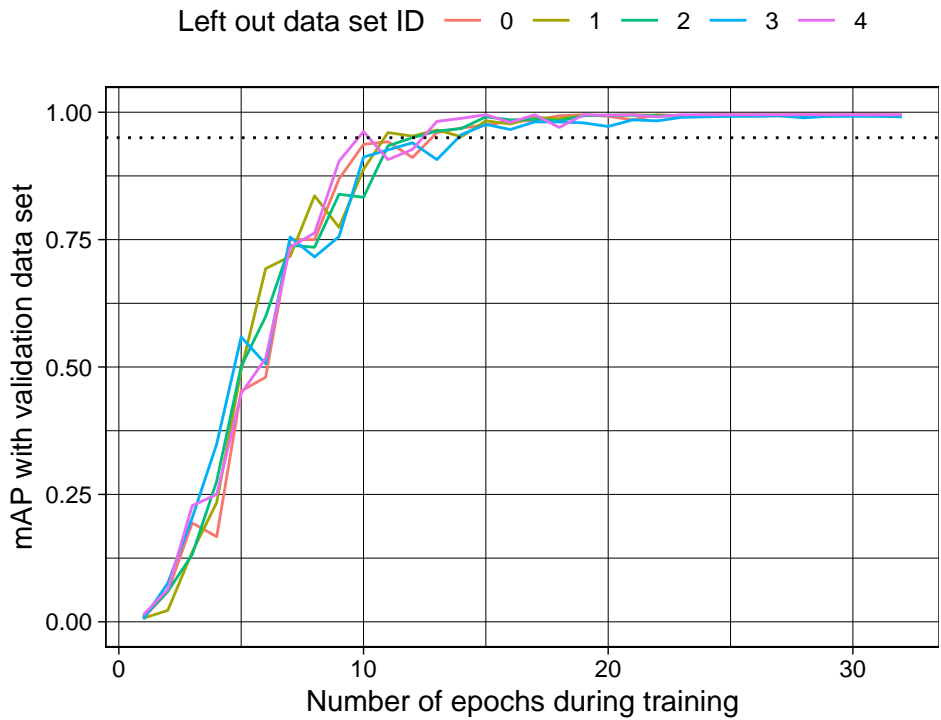

*Figure 4.6: Evolution of mAP calculated using test data set as function of number of epochs during training with 128 radiographs in the training data set. Dotted horizontal line indicate 0.95.*

## Metrics per Bone

The best model had it's metrics further analysed regarding bone class and all metrics were near 1 (Table 4.2).

*Table 4.2: Metrics per bone for best model.*

| Bone | Precision | Recall | mAP |
|---|---|---|---|
| radius | 0.997 | 1.000 | 0.995 |
| acessory carpal | 1.000 | 0.990 | 0.995 |
| proximal carpal row | 0.996 | 1.000 | 0.995 |
| distal carpal row | 0.996 | 1.000 | 0.995 |
| first carpal bone | 0.925 | 1.000 | 0.972 |
| metacarpal | 0.997 | 0.985 | 0.988 |

Detection of the first carpal bone required more epochs than other bones (Figure 4.7).

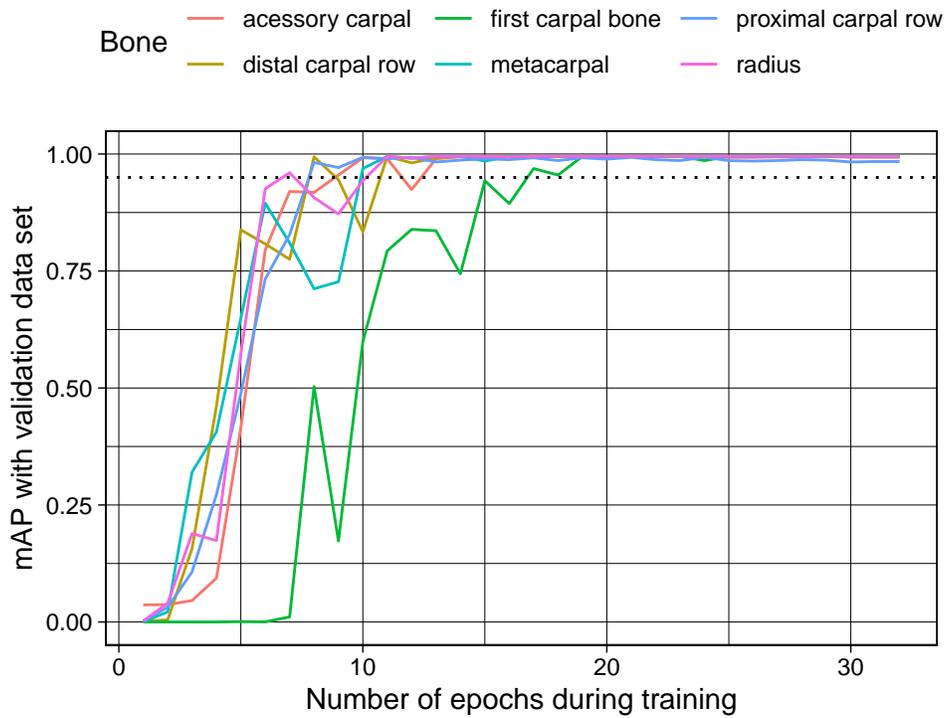

*Figure 4.7: Evolution of mAP of the best model. Dotted horizontal line indicate 0.95.*

### Inspection

The best model was used to generate a visual representation of the bounding boxes around the bones the model was able to detect (Figure 4.8).

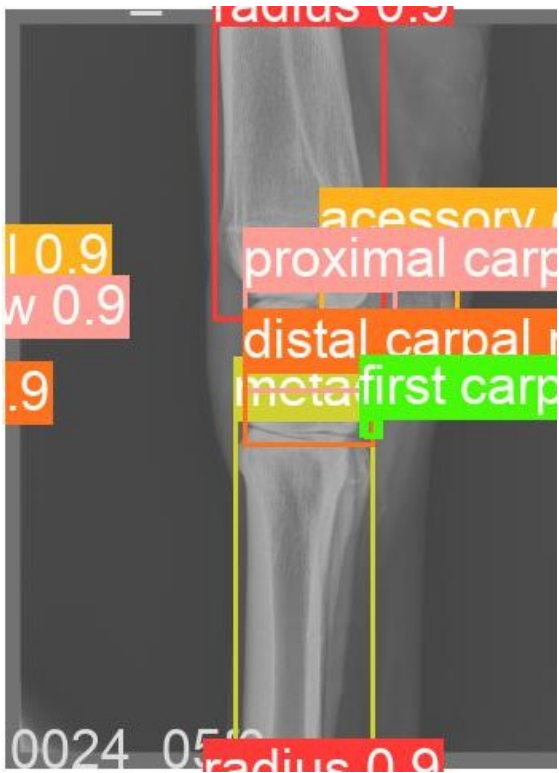

*With first carpal bone.*

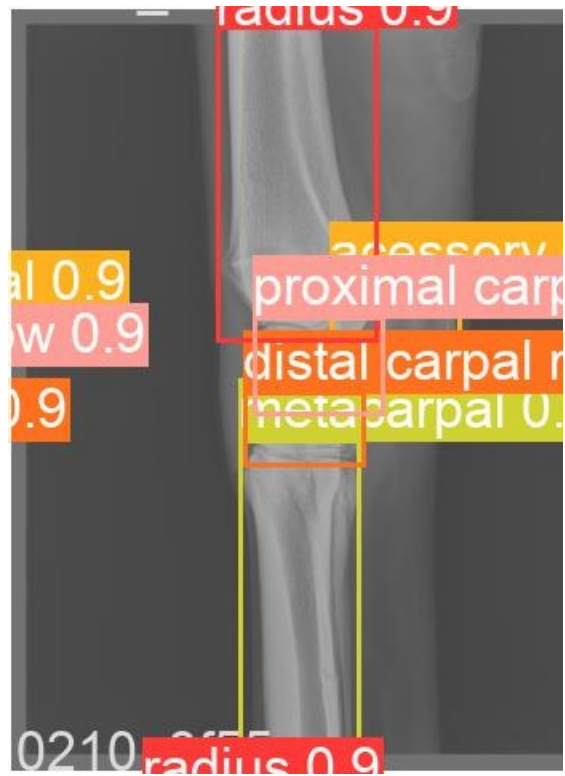

*Without first carpal bone.*

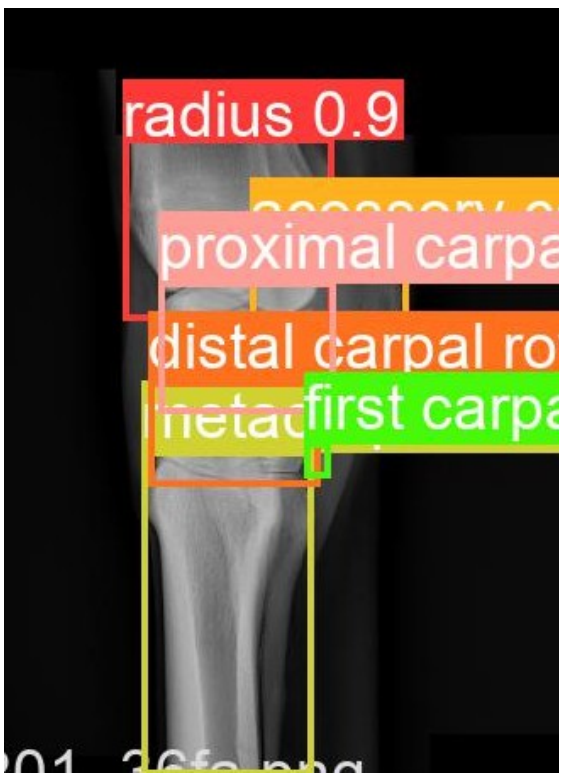

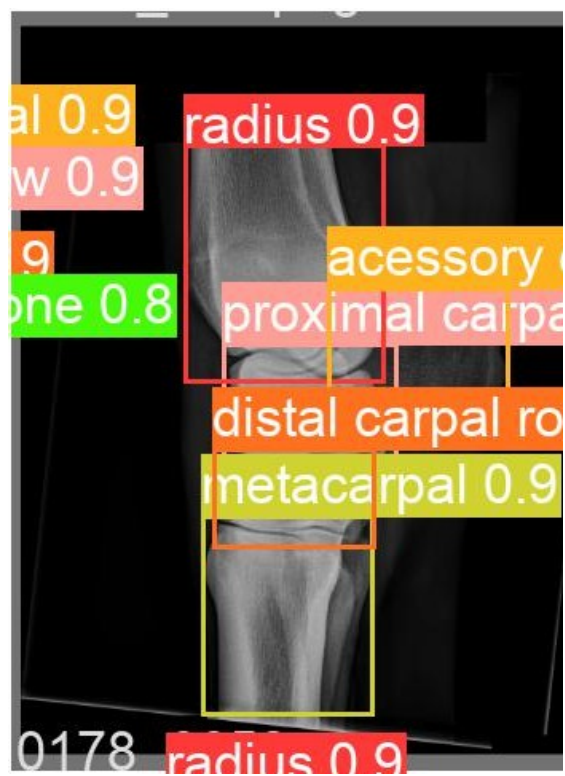

*With first carpal bone and text burned in cover by black rectangle.*   *Without first carpal bone and text burned in cover by black rectangle.*

*Figure 4.8: Illustration of detection as bounding boxes of left carpus dorsal 75º medial to palmarolateral oblique radiographs.*

# Discussion

We re-trained the YOLOv5s model[25] pre-trained on the COCO data set[31] for object detection of 6 bones (radius, acessory carpal, proximal carpal row, distal carpal row, first carpal bone, and metacarpal) in carpus DMPLO radiographs using different sizes of training data sets. For each different size of training data set, we conducted 5 experiments using leave-one-out cross-validation. After training, the model of each experiment was tested using a new data set not seen before by the models and metrics (precision, recall and mAP) were collected. Models trained with with 96 or more radiographs surprisingly achieved excellent metrics (above 0.95), including for the first carpal bone that has a average bounding boxes size of 0.1% of the radiograph. The best model (left out data set ID = 1 with 240 radiographs) achieved precision of 0.985, recall of 0.996 and mAP of 0.99.

Regarding the evolution of metrics with the number of epochs, metrics improved during the first 20 epochs when they reach 0.98 or higher and maintained stable in epochs higher than 20. Looking the evolution of metrics with the numbers of epochs of each bone for the best model, the training to detect the first carpal bone required the double of epochs than the other bones because YOLO based models "struggles to precisely localize some objects, especially small ones"[24]. The best model achieved near perfect (0.99 or above) detection of the first carpal bone with 20 epochs.

The best model has near perfect (0.99 or above) precision when detecting all bones except the first carpal bone that has good (0.90 or above) precision. We expected the first carpal bone to have a lower precision compare to other bones given it's smaller size (less than 0.1% of the radiograph) and we were surprise with the model high accuracy for the first carpal bone.

Our study has limitations. First, we only used a single radiographic projection of the carpus (dorsal 75º medial to palmarolateral oblique). We had other projections (dorsopalmar, dorsal 55º lateral to palmaromedial oblique, flexed lateromedial, and flexed dorsal 60º proximal dorsodistal oblique) and they were not used in this study because the first carpal bone was harder to detect given bones superimposed or juxtaposed. Given the surprising result with the first carpal bone, further study should be conducted using the other projections. Second, we only used radiographs provided by a single institution. We mitigated this by using leave-one-out cross-validation and a test set never seen by the model during training. By using a model available under a open source license and a small number of horses (198), we facilitated replication of this study in other regions. Last, we only used YOLOv5s as the start point of transfer learning. We considered the "small" member of the YOLOv5 architecture family suitable for our preliminary study with the objective of have a baseline to guide other authors regarding the smallest training data set they should use. Future studies should include larger members of the the YOLOv5 architecture family that use $1280 \times 1280$ pixels image as input.

The main goal of this study was to seek the answer to "what is the smallest training data set of a single radiograph view from a homogeneous population required to produce an object detection model with average accuracy above 0.95?". Our results suggest that studies of similar radiographs can achieve a high level of accuracy with only 100 images.

# Acknowledgements

We thank Dr. Anil Prabhu, Vincent Y. T. Tang, and Voss Y. T. Yu from the Hong Kong Jockey Club for provide access to the pre-import radiograph sets that were used to create the dataset used in this work.

## Funding
- City University of Hong Kong Postgraduate Studentship (by UGC-related research projects) 9610446

# Author Contributions
- Guarantors of integrity of entire study
    - Raniere Gaia Costa da Silva
- study concepts/study design
    - Michael Doube
    - Raniere Gaia Costa da Silva
- data acquisition
    - Raniere Gaia Costa da Silva
- data analysis/interpretation
    - Raniere Gaia Costa da Silva
- manuscript drafting or manuscript revision for important intellectual content
    - Raniere Gaia Costa da Silva
    - Michael Doube
- manuscript revision for important intellectual content

- – all authors
- approval of final version of submitted manuscript
    - – all authors
- agrees to ensure any questions related to the work are appropriately resolved
    - – all authors
- literature research
    - – Raniere Gaia Costa da Silva
    - – Michael Doube
- experimental studies
    - – Raniere Gaia Costa da Silva
- statistical analysis
    - – Raniere Gaia Costa da Silva
- manuscript editing
    - – all authors

# Conflicts of Interest

The authors declare they have no conflicts of interest.